\documentstyle[twoside,fleqn,espcrc2]{article}

\newcommand{\AmS}{{\protect\the\textfont2
  A\kern-.1667em\lower.5ex\hbox{M}\kern-.125emS}}

\title{Analytic Variational Investigation of Euclidean SU(3) Gauge Theory}

\author{N.D. Hari Dass and G.Subramoniam\address{Institute of Mathematical
Sciences , C.I.T Campus, Madras 600 113, India}%
\thanks{\ This work was carried out under the project  on Lattice
Gauge Theories (SP/S2/K22/87) sponsored by DST ( Department of Science and
Technology, India).}}

\begin{document}

\begin{abstract}
Analytic variational techniques for lattice gauge theories based on the
Rayleigh-Ritz({\bf RR}) method were previously developed for euclidean SU(2)
gauge theories in 3 and 4 dimensions. Their extensions to SU(3) gauge theory
including applications to correlation functions and mass gaps are presented
here.
\end{abstract}

\maketitle

\section{Introduction}

Despite impressive strides made in the Monte Carlo(MC) studies of lattice
gauge theories,  a satisfactory understanding of the wavefunctionals of
these theories is still lacking. Motivated by the success of variational
methods of quantum mechanics in addressing such issues,  a program of
developing these methods for lattice systems was undertaken. At first these
methods were applied to spin systems, both discrete and continuous, in two
dimensions, with success. Subsequently, the much harder problem of applying
these methods to 4-dimensional gauge theories was undertaken and successfully
solved for the case of U(1) and SU(2) gauge theories. Details of these and
other related works along with the relevant references can be obtained from
\cite {Hari84}.

The basic idea is to apply the {\bf RR} technique to the spectrum of the
transfer matrix, denoted symbolically by {\bf T}. One makes an ansatz for
the largest eigenvalue state guided by the usual intutions of nodelessness,
symmetry etc.

The simplest such ansatz for a lattice gauge theory is
\begin{equation}
\psi = exp(\alpha \sum Re tr P_i)
\end{equation}
where $P_i$ are the relevant plaquette variables lying on a ``time'' slice
with tr being taken over some irr.\ of the group, and $\alpha$ a variational
parameter. Recall that {\bf T} connects configurations on neighbouring time
slices. The estimate for the largest eigenvalue is then given by
\begin{equation}
\Lambda_0 = max_{\alpha} (\psi,{\bf T}\psi)/(\psi,\psi)
\end{equation}
The denominator in (2) is the partition function of a 3-dimensional theory
with coupling $2\alpha$, while the numerator can be interpreted as the
partition
function of a generalised 3-dimensional theory where the links per site are
doubled and with anisotropic couplings $\alpha + \beta/2$, $\beta$.

Our idea was to construct the appropriate transfer matrices for these lower
dimensional problems and then apply the {\bf RR} method to their spectra, and
so on. Under the simplifying assumption justified \'{a} posteriori, that the
same variational parameters are used for all the subsequent ans\"{a}tze, one
arrives at  the variational estimate for the free energy per
site of the form
\begin{equation}
-{\bf F}_4 = max_{\alpha}\left[{\bf F}^{H}(\beta_H)+G(\alpha)\right]
\end{equation}
where ${\bf F}^{H}$ is the free energy for a 4-dimensional unit hypercube with
coupling $\beta_H=3/2\alpha +\beta/4$, while $G(\alpha)$ is expressible in
terms of the
norms of the various ans\"{a}tze (details are given in \cite {Hari84}). The
unit hypercube partition function as well as $G(\alpha)$ are evaluated by
making
use of group character expansions
\begin{equation}
exp (x Retr_f (g)) = \sum d_r tr_r(g) b_r(x)
\end{equation}
where $tr_i(g)$ is the character in the  i-th irrep.\  whose dimension is
$d_i$.
 For SU(2) $b_r(x)=I_{2r+1}(2x)/x $ where the $I_r(x) $ is a
modified Bessel function. For SU(3) their analytic form is not known. $f$
corresponds to the fundamental rep.

The main result which made the analytic investigation of the SU(2) case
possible
is the explicit expression for the partition function of a unit hypercube in
terms of a product of 16 SU(2) 6-j symbols \cite {Hari84a}. This result was
obtained by generalising the graphical rules of Yutsis, Vanagas and Levinson
\cite {Yuts68}. Sybolically expressed, this result reads
\begin{equation}
e^{-{\bf F}^{H}}=\sum d_{r1}b_{r1}....d_{r24}b_{r24}\prod  6j's
\end{equation}
For SU(3), however, there are additional complications due to i)the complex
nature of irreps, ii)multiplicity problems associated with weights and tensor
decompositions. These are being investigated at present. However, substantial
contribution to (5) comes from surfaces that only depend on $d_i$
and the number of singlets $N_{rst}$ contained in the tensor
product of 3 irreps. Even though the latter is affected by multiplicity
problems too, a programme was developed based on the elegant prescription given
in  \cite {Gadi91}.

In practice, the sum (5) is evaluated by finding the contribution of all
possible closed surfaces that can be embedded in the unit hypercube.
A complete classification of such surfaces is given in
\cite {Hari84a}. For example, each surface of area A and topology of a sphere
contributes $\sum d_r^2 b_r^A$ to (5), while each surface with area A and
torus topology contributes $\sum b_r^A$. There are, for example, 8 spheres
with A=6 and 3 torii with A=16 in the unit hypercube. Likewise, surfaces
obtained
by gluing 3 disks of area $A_r$ spanned by plaquettes in the irrep r, $A_t$ by
plaquettes in the irrep t and $A_s$ in irrep s, contribute
$\sum N_{rst}d_rd_td_s b_r^{A_r}b_s^{A_s}b_t^{A_t}$.
 More complicated surfaces require
a knowledge of the 6-j symbols. The sum in the above formulae(except(5)) is
only
over nontrivial irreps.

The $b_r(x)$ for $SU(3)$ were calculated by generating their series expansions
upto $x^{100}$ using an algorithm to find the number of times
the irrep $r$ occurred in $(f\oplus \bar f)^n$ by a recursion relation. The
resulting $b_r$'s were checked against a numerical evaluation of their integral
representation for $x$ as large as 10, and agreement was seen to double
precision accuracy.A weak coupling expansion was also developed using the
symbolic manipulation package FORM and again good agreement was seen with the
series expansion.

At the moment we have restricted the
sum(5) to only those surfaces for which one at most needs the $N_{rst}$. It is
encouraging(also from the point of more accurate extensions of this method)
that in the range $0 < \beta <2.25 $(this corresponds to the range $(0, 6.75)$
for the coupling used in literature) this truncation works quite well as will
be shown by the results.

\section{Correlation Functions And Mass Gaps}

There are three classes of plaquette-plaquette correlations within the unit
hypercube that are relevant for the study of mass-gaps;these are a)when the
plaquettes on both the time-slice are in the same spatial plane and b)when
they are in planes orthogonal to one another(this class will be called dh3).
Case a) further splits into a1)when the two plaquettes can be embedded in a
cube(dh2) and a2) otherwise(dh1). It is useful to factor out $b_0(x)^{24}$ in
(5)
(here $0$ refers to the trivial irrep) and write
\begin{equation}
-{\bf F}^H=24 ln b_0(x) + ln (1+A)
\end{equation}
where $A$ is the sum of contributions from various surfaces, examples of some
having been given before, except that $b_r$'s in those expressions should be
replaced by $B_r=b_r/b_0$. The average plaquette(normalised) is then given by
\begin{equation}
\langle P \rangle =B_f(\beta_H) +1/72 A' (1+A)^{-1}
\end{equation}
The connected correlation function between two given plaquettes $P_1$ and $P_2$
can be obtained as follows: let the coupling for these two be $\beta_1$ and
$\beta_2$ with all other couplings being $\beta_H$. The free energy in that
case
is given by
\begin{equation}
e^{-{\bf F}^H}={ b_0(\beta_H)}^{22} b_0(\beta_1)
                                    b_0(\beta_2)(1+A(\beta_1,\beta_2))
\end{equation}
Here $A$ contains contributions only from surfaces in which both the plaquettes
$P_1$ and $P_2$ are embedded.
The connected correlation function(normalised) is given by
\begin{equation}
\langle P_1 P_2\rangle =1/9(\frac{\partial}{\partial\beta_1} \frac{\partial}
{\partial\beta_2})(-{\bf F}^H)|_{\beta_1=\beta_2=\beta_H}
\end{equation}
It is clear
from the above that a lot fewer surfaces contribute to the connected
correlation
lengths than to $\langle P\rangle$.
At low
$\beta_H$, the dominant contribution to the average plaquette is
$B_f(\beta_H)$,
and all connected correlation functions nearly vanish. Only
dh2 receives contribution from one out of the 8 spheres of
area 6. With increasing $\beta_H$ other surfaces start contributing and all
the three classes become appreciable.

\subsection{Mass Gap estimates}

Having obtained a variational ``ground'' state $|0\rangle^t$ for the transfer
matrix, we can extend the technique to obtain estimates for the lower
eigenvalues and hence the mass gaps. There are some important technical
differences in
implementing this strategy for lattice systems as compared to quantum
mechanics.
We  summarise these points
and present the details elsewhere:
a) in the variational treatment of quantum mechanical systems, the ans\"{a}tze
for the excited states can be chosen quite independently of the ground state
ansatz. But in the RR applied to lattice theories the trial excited state
functionals have to be of the form $F|0\rangle^t$ where $F$ is a polynomial
functional;otherwise, the mass gap grows with volume in the thermodynamic
limit.

b) if the excited state `projector' F is expressed as $F=\sum c_r \theta_r$
with
$\theta_r$ a basis for such projectors, a necessary condition  for the problem
determining the eigenvalues of the transfer matrix to be well defined is
\begin{equation}
\frac{\langle 0|{\bf T}\theta_r| 0\rangle^t}{\langle 0|{\bf T}|0\rangle^t}=
\frac {\langle 0|\theta_r|0\rangle^t}{\langle 0|0\rangle^t}
\end{equation}
Otherwise, the  problem of determining  the eigenvalues of
$ND^{-1}$ where the matrices $N$ and $D$ are given by $N_{rs}=\frac{\langle
0|{\bf T}\theta_r\theta_s|0\rangle^t}{\langle 0|{\bf T}|0\rangle^t}$ and
$D_{rs}=\frac{\langle 0|\theta_r\theta_s|0\rangle^t}{\langle 0|0\rangle^t}$,
becomes ill-defined  in the sense that while $D_{rs}$
grows as $volume$ in the thermodynamic limit, $N_{rs}$ grows as $vol ^2$.

Situations where (10) is satisfied automatically are
i) due to symmetry reasons both $\langle 0|{\bf T}\theta_r|0\rangle^t$ and
$\langle 0|\theta_r|0\rangle^t$ vanish;this happens, for example, in the X-Y
model when $\theta_r$ are rotationally non-invariant.
ii) when $|0\rangle^t$ is an exact eigenstate of ${\bf T}$.
iii) when $|0\rangle^t$ is of the form $exp(\sum \mu_r\theta_r)$ where
$(\mu_r)$
are determined variationally, in which case (10) are  the
conditions for the stationarity of $\Lambda_0$ w.r.t $(\mu_r)$.

In view of the last remark we should build the excited state projectors
only out of $P$'s. Nevertheless, in the case of $SU(2)$, Wilson loops of
length 6
could also be considered as it was seen that in the range of $\beta$-values of
interest, (10) was satisfied quite well. But $D_{rs}$
in that case required the knowledge of 6-j symbols; hence for $SU(3)$ we have
restricted our attention to only plaquettes in the
fundamental rep. Following standard procedure, one then constructs linear
combinations of these that transform irreducibly under the cubic group. With
this choice, $N$ and $D$ are already diagonal. To the accuracy considered the
various cubic irreps were nearly degenerate in mass(which is not the true
situation) and hence only the lowest (scalar)mass estimates will
be shown in the table.

\section{Results }

In the strong  and weak coupling ends the analysis can be carried out
to yield $\alpha$ in analytic form. In the intermediate regions the analysis
has to be made numerically but without MC.

The first four terms of the exact free energy in both these ends are reproduced
for all gauge groups and dimensions(space-time);actually for d=4 the coefft of
the fourth term in the weak coupling is numerically very close, but not exactly
equal to, the exact result. The variational parameter $\alpha=\beta/2+c\beta^5$
for small $\beta$ and equals $7/9\beta$ for large $\beta$.

In the case of $SU(2)$ the average plaquette for d=3 was indistinguishable
from high statistics MC; in d=4 it agreed to within
.1\% everywhere except the
crossover region where the maximum discrepancy was about 2\%. The specific heat
in d=4 also agreed very well. The string tension and mass gap agreed very well
upto the crossover but beyond there was a noticeable deterioration. For further
details see \cite{Hari84}. In the mass gap estimates inclusion of operators in
higher irreps. \ made practically no difference while there was a dramatic
decrease in the mass gap when the effective dimensionality of the transfer
matrices was increased. Also inclusion of larger Wilson loops improved the
estimates.

\begin{table}[t]
\caption{SU(3): Local Quantities}
\begin{tabular}{rrrrr}
\hline
$\beta_H$  & $\beta$  &$\langle P\rangle$&$\langle P\rangle_4$&$\langle P
\rangle_{an}$ \\
\hline
$ 0.6 $    & $ .6000$ & $ .1138$ & $ .1168$ & $ .1164$ \\
$ 1.0 $    & $ .9950$ & $ .2028$ & $ .2078$ & $ .2040$ \\
$ 1.2 $    & $1.1880$ & $ .2501$ & $ .2566$ & $ .2517$ \\
$ 1.4 $    & $1.3720$ & $ .3012$ & $ .3017$ & $ .3012$ \\
$ 1.6 $    & $1.5410$ & $ .3512$ & $ .3592$ & $ .3522$ \\
$ 1.8 $    & $1.6960$ & $ .4019$ & $ .4330$ & $ .4044$ \\
$ 2.0 $    & $1.8500$ & $ .4586$ & $ .4821$ & $ .4554$ \\
\hline
\end{tabular}
\end{table}

In d=3 our variational results for SU(3) looked qualitatively very much like
the results for SU(2) in d=3; there was a noticeable peak in the specific heat
curve which however was not as sharp as the one in d=4.Here we did not compare
the results with MC data.In the weak coupling region analytical results were
obtained with the help of FORM and the first four terms of the free energy
density agreed with exact results. The details of this case have not been
given here but will be reported elsewhere. Efforts are under way to enlarge the
scope of these calculations by considering more complicated ans\"{a}tze.
Inclusion of fermions is also being considered.

The results for $SU(3)$ in d=4 are summarised in two tables. A two-fold check
with
MC was performed. The first of these consisted in simulating the unit hypercube
itself;the `infinite' lattice was simulated on a $4^4$ lattice. One noticed
that
when $\beta^H$ was small there was no appreciable difference between $\langle
 P\rangle$ obtained by the two simulations. But as $\beta$ increased there was
a
considerable difference between them;but when the $\langle P\rangle$ of the
hypercube simulation at $\beta_H$ was compared with that of the 4-site
simulation at
$\beta$, there was very good agreement. This can be viewed as a
MC realisation of the method described here. On the other hand, since in our
$SU(3)$ studies we have not been able to include all the possible surfaces in
(5), the MC of the hypercube served as a check on the accuracy of the
analytical
results represented by $\langle P\rangle_{an}$ in table 1 and by dh2(an) in
table 2.
The hypercube MC was performed with 50,000 sweeps each. The tables show
that as far as $\langle P\rangle$ is concerned, upto $\beta\sim 2.25$ the
RR estimate is very good. Comparison of the connected correlations is trickier;
at low $\beta$ these are very very small and MC is unable to measure them
reliably. At larger $\beta$, however, the error in $\langle P \rangle_{an}$
becomes significant due to the neglect of many surfaces in (5). Thus reliable
comparison can be made only in the window (1.2-2.2) which fortunately covers
the interesting range.But even in this range only dh2 could be sensibly
compared.

\subsection{Future Prospects}
 From the behaviour of the mass gap in both $SU(2)$ and $SU(3)$ cases it is
clear that the ansatz (1) works very well for local quantities like
$\langle P\rangle$ but in the region close to the crossover it is not so
efficient in describing long distance correlations. On the other hand
this was the simplest ansatz one could have considered and it is surprising
that it worked as well as it did. It is clear that larger operators have to be
included;also the weakest link in the present implementation is at the stage
where the same type of ansatz was used for the 2-dim transfer matrix. But in
that
case the ansatz predicts strictly vanishing d=2 correlations(connected). Any
further improvement must circumvent this. An effective way of introducing
larger
operators is to choose ${\bf T}^n|0\rangle^t$ as the trial ground state. This
will then involve the enumeration of surfaces embedded in larger hypercubes.
This
is under study.
\begin{table}[t]
\caption{SU(3): Correlations and Masses}
\begin{tabular}{rrrr}
\hline
  $\beta_H$ &   dh2    &   dh2(an)&   mass    \\
\hline
  $0.8$     &-.00002   & .00006   &           \\
  $1.0$     & .00005   & .00018   &  5.5      \\
  $1.2$     & .00046   & .00041   &  4.4      \\
  $1.4$     & .00076   & .00078   &  4.6      \\
  $1.6$     & .00132   & .00121   &  3.6      \\
  $1.8$     & .00198   & .00148   &  4.0      \\
  $2.0$     & .00237   & .00129   &  4.2      \\
\hline
\end{tabular}
\end{table}

\end{document}